\documentclass[journal,draftcls,onecolumn,12pt]{IEEEtran}

\usepackage{cite}   
\usepackage{amsmath,amssymb}
\usepackage{enumerate}
\usepackage{array}
\newtheorem{definition}{Definition}
\newtheorem{theorem}{Theorem}
\newtheorem{corollary}{Corollary}
\newtheorem{proposition}{Proposition}

\usepackage{graphicx}
\usepackage{epsfig}

\begin{document}

\title{On Maximum Contention-Free Interleavers and Permutation 
  Polynomials over Integer Rings}  

\author{Oscar Y. Takeshita  \\
Dept. of Electrical and Computer Engineering \\
2015 Neil Avenue \\
The Ohio State University \\
Columbus, OH 43210 \\ 
Takeshita.3@osu.edu\\
\vspace{2em}
Submitted as a Correspondence to the IEEE Transactions on Information Theory\\
April 28, 2005\\
Revised November 1, 2005
}
\date{\today}

\maketitle

\begin{abstract}
  An interleaver is a critical component for the channel coding
  performance of turbo codes.  Algebraic constructions are of
  particular interest because they admit analytical designs and
  simple, practical hardware implementation. Contention-free
  interleavers have been recently shown to be
  suitable for parallel decoding of turbo codes. In
  this correspondence, it is shown that permutation
  polynomials generate maximum contention-free interleavers, i.e.,
  every factor of the interleaver length becomes a possible degree
  of parallel processing of the decoder. Further, it is shown by computer simulations
  that turbo codes using these interleavers perform very well for the
   3rd Generation Partnership Project (3GPP) standard. 
\end{abstract}

\begin{keywords}
Turbo code, interleaver, permutation polynomial, contention-free,
algebraic, quadratic, parallel processing.  
\end{keywords}

\pagebreak

\section{Introduction}
Interleavers for turbo codes~\cite{sun:tak:pp, nim:isit04, berrou:int,
  Bravo, giulietti, Sadjadpour, tak:cos:it, crozier1,
  danesh:it,tak:cos:linear, dolinar, Berrou} have 
been extensively investigated.  Recently, Sun and Takeshita~\cite{sun:tak:pp}
suggested the use of permutation polynomial-based interleavers over
integer rings. In particular, quadratic polynomials were emphasized;
this quadratic construction is markedly different from and superior to
the one proposed earlier by Takeshita and
Costello~\cite{tak:cos:it}\footnote{The construction
  in~\cite{tak:cos:it} generates interleavers typically with the
  performance and statistics of a randomly generated 
  interleaver but with the advantage of a very simple generation.} in turbo
coding applications. The algebraic approach in~\cite{sun:tak:pp} was
shown to admit analytical design of an interleaver matched to the
constituent convolutional codes. The resulting performance was shown
to be better than S-random interleavers~\cite{dolinar} for relatively
short block  
lengths and parallel concatenated turbo codes; we show in this
correspondence that even for moderate block lengths (4096 information
bits) an excellent performance can be obtained. An iterative
turbo decoder needs both an interleaver and a deinterleaver. Ryu and
Takeshita have also shown a necessary and sufficient condition for a
quadratic permutation polynomial to admit a quadratic
inverse~\cite{ryu:tak:qinv}. Moreover, the simplicity of the algebraic
construction in~\cite{sun:tak:pp} implies efficient implementations as
one witnesses in~\cite{jpl2}.

The decoding of turbo codes is performed by an iterative process in
which the so-called extrinsic information is exchanged between
sub-blocks\footnote{There are typically two or more sub-blocks in an
iterative turbo decoder, each implementing a soft-input soft-output
decoding algorithm of a convolutional code.} of the iterative
decoder. The parallel processing of iterative decoding of turbo
codes is of interest for high-speed decoders. Aspects of implementations
of parallel decoders in chips and expected performance are studied
in~\cite{dobkin}. Interleaving of
extrinsic information is one important aspect to be addressed in
parallel decoders because a memory access contention, as explained in
this section, may appear during the exchange of extrinsic information
between the sub-blocks of the iterative decoder~\cite{giulietti}. The
first approaches to solve the memory access contention problem simply
avoided it by constraining the interleavers to be contention-free as
in~\cite{giulietti,Blankenship,nim:isit04,berrou:int}. For these 
type of constrained constructions of interleavers, Nimbalker {\em et
  al.} have shown that only a very small fraction of all interleavers
are suitable for parallel processing of iterative
decoding~\cite{nim:isit04}. They have also proposed a new 
construction of a modified dithered relatively prime\footnote{The DRP
interleaver construction is one of the best known for turbo codes
with excellent error rate performance.} interleaver~\cite{crozier1}
(DRP) interleaver. If the interleaver is required to be left
unconstrained (e.g., the interleaver cannot be modified because it is
already part of a standard), then the memory contention problem can
still be solved as shown in~\cite{tarable,thul} but at a cost of
additional complexity.

\begin{figure}[htbp]
\centering
\includegraphics{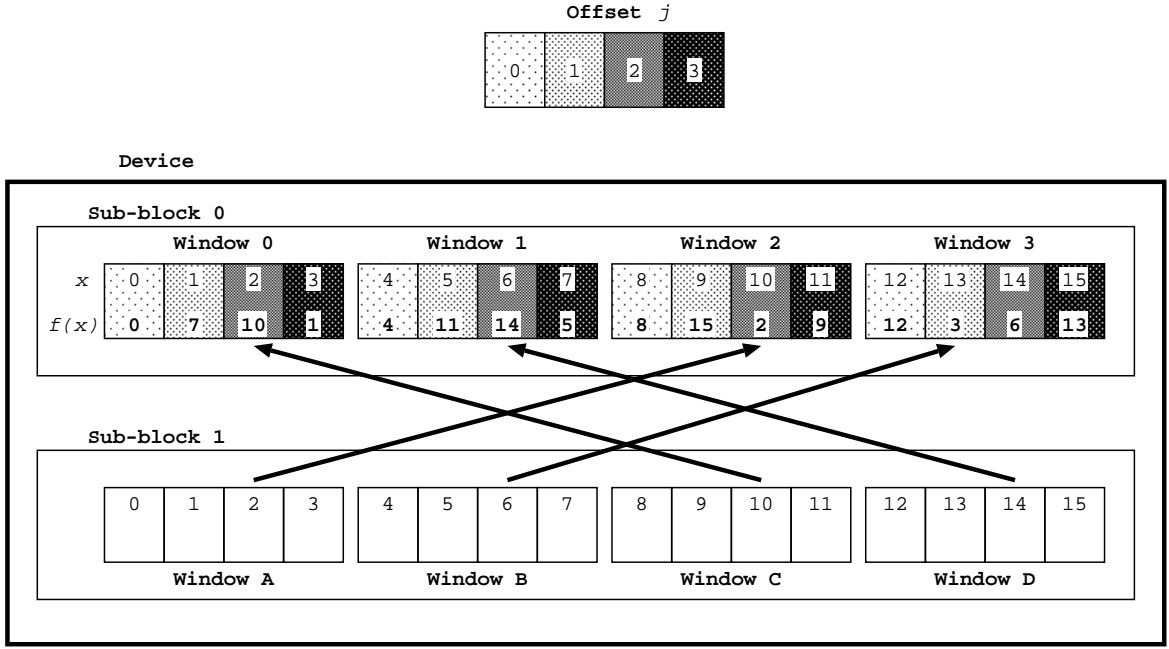}
\caption{Example of a contention-free property for $W=4$, $M=4$,
  and $N=16$.} 
\label{fig:memory}
\end{figure}

In this correspondence, we approach the memory contention problem
using constrained interleavers that are contention-free. The advantages
of our approach are its low complexity induced from an 
algebraic solution but with no apparent error rate performance
degradations against any good interleavers. The 
contention-free condition is illustrated in Fig.~\ref{fig:memory}
through an arbitrary device (not necessarily a turbo decoder). The
device has two sub-blocks. Each of the $N=16$ cells in sub-block 0
needs to fetch data in a one-to-one fashion from the $N=16$ cells in
sub-block 1. If sub-block 0 processes data in a serial fashion, and
its cells fetch data sequentially from left to right $(x=0,x=1,\ldots x=15)$
then the sequence $(f(0)=0,f(1)=7,\ldots, f(15)=13)$ indicates the
addresses of the cells in sub-block 1 from which 
data is extracted. The function $f(x)$ describes the interleaver. If
sub-block 0 
processes data in a parallel fashion using $M=4$ processors, then
sub-block 0 is split in windows of size $W=4$. A cell in a window has
an offset value $0\leq j<W$  (different values of offsets are
shown in different shades). Each of the four processors fetches data
simultaneously always at a particular offset $j$. The contention-free property
requires that for a fixed offset, exactly one cell is accessed from
each of the four windows in sub-block 1. An example is shown in
Fig.~\ref{fig:memory} for the offset $j=2$. This implies that if cells
in sub-block 1 are 
organized in four independent memory units for each of the windows A,
B, C and D then we do need to worry about memory contention, i.e., two or
more processors in sub-block 0 trying to simultaneously fetch data in
the same memory unit in sub-block 1.

In turbo coding applications, this property is also desirable in the
reverse order, i.e., when sub-block 0 and 1 switch roles. A
mathematical description of the contention-free condition
from~\cite{nim:isit04} is now given.  
The exchange and processing of a sequence of $N=MW$ extrinsic information
symbols between sub-blocks of the iterative decoder can be
parallelized by $M$ processors working on window sizes of length $W$
in each sub-block without contending for memory access provided that
the following condition holds for both the
interleaver $f(x)$, $0\leq x <N$ and deinterleaver $g(x)=f^{-1}(x)$:

\begin{equation}
\lfloor \pi(j+tW)/W\rfloor \not = \lfloor\pi (j+vW)/W\rfloor
\label{eq:thecondition}
\end{equation}

where $0\leq j<W$, $0\leq t<v<N/W$, and $\pi(\cdot)$ is either
$f(\cdot)$ or $g(\cdot)$. 

If an interleaver is {\em contention-free} for
all window sizes $W$ dividing the interleaver length $N$, it
will be called a {\em maximum contention-free} interleaver. 
We show in this correspondence that permutation polynomials
over integer rings {\em always} generate {\em maximum contention-free}
interleavers. 

This correspondence is organized as follows. In section II, we 
review a result for quadratic permutation polynomials\cite{sun:tak:pp, ryu:tak:qinv}
over the integer ring $\mathbb{Z}_N$ and an elementary number theory
proposition~\cite{Hardy} needed for the main theorem. The main result
is derived in section III, and examples and computer simulation
results are given in section IV. Finally, conclusions are discussed in
section V.   

\section{Quadratic Permutation Polynomials over Integer Rings}
In this section, we establish notation, restate the criterion for existence of quadratic permutation 
polynomials over integer rings and, restate a result in number theory. The interested reader is referred
to~\cite{sun:tak:pp,ryu:tak:qinv} for further details. Given an integer $N \geq 2$, a
polynomial\footnote{It can be shown 
  that the exclusion of a constant coefficient $f_0$ in $f(x)$ does not make this problem less
  general.} $f(x) = f_1 x + f_2 x^2 \pmod{N}$, where  
$f_1$ and $f_2$  are non-negative integers, is said to be a
quadratic permutation polynomial over the ring of integers $\mathbb{Z}_N$
when $f(x)$ permutes $\{0, 1, 2, \ldots, N-1 \}$\cite{sun:tak:pp, Mullen1}.

In this correspondence, let the set of primes be $\mathcal{P} = \{2,3,5,7,\ldots   \}$. 
Then an integer $N$ can be factored as $N = \prod\nolimits_{p \in
  \mathcal{P}}  p^{n_{N,p}}$, where $n_{N,p} \geq 1$ for a finite number of
$p$'s and $n_{N,p}=0$ otherwise. For example, if $N=3888=2^4\times 3^5$ we
have $n_{3888,2}=4$ and $n_{3888,3}=5$.  
For a quadratic polynomial $f(x)=f_1 x + f_2 x^2 \pmod{N}$, we will abuse the
previous notation by writing $f_2 = \prod\nolimits_{p \in \mathcal{P}}
p^{n_{F,p}} $, i.e., the exponents of the 
prime factors of $f_2$ will be written as $n_{F,p}$ instead of the more
cumbersome $n_{{f_2},p}$ because we will be mainly interested in the
factorization of the second degree coefficient. \\
Let us denote  $a$ divides $b$ by $a|b$ and by $a\nmid b$
otherwise. The greatest common divisor of $a$ and $b$ is denoted by
$\gcd(a,b)$. The necessary and sufficient
condition for a quadratic polynomial $f(x)$ to be a permutation
polynomial is given in the following proposition.  

\begin{proposition}\cite{ryu:tak:qinv}\cite{sun:tak:pp}
 Let $N = \prod\nolimits_{p \in \mathcal{P}}  p^{n_{N,p}} $.
The necessary and sufficient condition for a quadratic polynomial $f(x) = f_1 x + f_2 x^2 \pmod{N}$ 
to be a permutation polynomial can be divided into two cases. 

\begin{enumerate}
\item Either $2 \nmid N$ or $4 |N$ (i.e., $n_{N,2}\not = 1$)\\
$\gcd(f_1, N)=1$ and $   f_2 = \prod\nolimits_{p \in \mathcal{P}}  p^{n_{F,p}}, n_{F,p} \geq 1  $, $\forall p$
such that  $n_{N,p} \geq 1$.

\item $2|N$ and $4\nmid N$ (i.e., $n_{N,2}=1$)\\
$f_1+f_2$ is odd, $\gcd(f_1,\frac{N}{2}) = 1$ and $f_2 = \prod\nolimits_{p \in \mathcal{P}}  p^{n_{F,p}}, n_{F,p} \geq 1$, $\forall i$ 
such that $p \neq 2$ and $n_{N,p} \geq 1$.
\end{enumerate}
\label{prop:pp2}
\end{proposition}

How many permutation polynomials are there? For
example, if the interleaver length is $N=256$ then we determine from
case 1) of Proposition~\ref{prop:pp2} that $f_1\in 
\{1,3,5,\ldots,255\}$ (set of numbers relatively prime to $N$) and
$f_2=\{2,4,6,\ldots,254\}$ (set of numbers that contains 2 as a
factor). This gives us $128\times127=16256$ possible pairs of
coefficients $f_1$ and $f_2$ that make $f(x)$ a permutation
polynomial; if $N$ is a power of two then there are approximately
$N^2/4$ possible pairs of coefficients. However, if $N$ is a prime
number then there are no polynomials of the form $f(x)$ for a non-zero
$f_2$. This may be perceived as a deficiency of the
construction because certain interleaver lengths must be
avoided. However, even restricting to powers of two gives plenty 
of possibilities and covers meaningful interleaver lengths. In
general, the number of permutation polynomials is not a smooth
function of $N$. 

Let us denote that $x$ is congruent to $y$ modulo $N$ by $x\equiv y
\pmod{N}$; this means that there exists an integer $k$ such that $x=y+kN$.
The following elementary number theory proposition is 
used for deriving the main theorem (Theorem~\ref{th:main}) of this
correspondence. 

\begin{proposition}\cite{Hardy}
Let $M$ be an integer. Suppose that $M | N$ and that $x \equiv y \pmod{N}$. Then $x \equiv y \pmod{M}$.
\label{prop:reduction}
\end{proposition}
The proof follows by noting that $x=y+kN=y+kWM$, where $W=N/M$.

\section{Maximum Contention-Free Permutation Polynomials Interleavers}

The following defines a contention-free interleaver on its maximum
extent. 

\begin{definition}
An interleaver is {\em maximum contention-free} ({\em MCF}) when the
interleaver is contention-free for every window size $W$ which is a factor
of the interleaver length $N$.
\end{definition}

It is natural that the previous definition implies that we potentially
have a degree of parallel processing of any
soft-input soft-output algorithm by $M=N/W$
processors, i.e., each factor of $N$ is a possible number of parallel
processors. We show that quadratic permutation polynomials always generate
interleavers that are {\em MCF}. 

\begin{theorem}
Let $f(x)=f_1x+f_2x^2 \pmod{N}$ be a quadratic permutation
polynomial. Then $f(x)$ generates a {\em MCF}
interleaver.
\label{th:main}
\end{theorem}

\begin{proof}
We first verify condition (\ref{eq:thecondition}) for $f(x)$ and then
for $g(x)=f^{-1}(x)$. 

Let 

\[
Q_t=\left\lfloor \frac{f(j+tW)}{W}\right\rfloor
\quad \mbox{and} \quad
Q_v=\left\lfloor \frac{f(j+vW)}{W}\right\rfloor
\]

then

\[
f(j+tW)=Q_tW+[f(j+tW) \pmod{W}] \quad \mbox{and} \quad f(j+vW)=Q_vW+[f(j+vW) \pmod{W}]
\]

We must show that $Q_t\neq Q_v$ for $t-v\not \equiv 0\pmod{M}$ and any $0\leq
j<W$.

Assume $Q_t=Q_v$. Then 

\begin{equation}
Q_t-Q_v=\frac{f(j+tW)-[f(j+tW) \pmod{W}]-f(j+vW)+[f(j+vW)
  \pmod{W}]}{W}=0
\label{eq:q1q2}
\end{equation}

Using Proposition~\ref{prop:reduction} and observing that 

\begin{equation}
f(j+tW) \equiv f_1j+f_2j^2\pmod {W} \quad \mbox{and} \quad f(j+vW)
\equiv f_1j+f_2j^2\pmod {W},
\label{eq:minipp}
\end{equation}

we conclude $[f(j+tW) \pmod{W}]=[f(j+vW) \pmod{W}]$ and therefore the
absolute value of equation
(\ref{eq:q1q2}) can be simplified as 

\begin{equation}
|Q_t-Q_v|=\frac{|f(j+tW)-f(j+vW)|}{W}=0
\label{eq:remmod2}
\end{equation}

By noting that $(j+tW) \neq (j+vW)$  and
that $f(x)$ is a permutation polynomial, we conclude
$f(j+tW)\neq f(j+vW)$ and we have a 
contradiction in (\ref{eq:remmod2}).

To verify condition (\ref{eq:thecondition}) for the inverse polynomial
$g(x)$, we start by observing that permutation polynomials form a
finite group $G$ under function composition, i.e., $f(f(x))$ is a
permutation polynomial and the inverse function can be found by a
sufficient number of function compositions of $f(x)$ to itself. In
group theory parlance, $f(x)$ generates the group
$G$. It now suffices to show that every element in $G$, which
includes the inverse function $g(x)$, satisfies
(\ref{eq:thecondition}). This is easily shown by realizing that 
(\ref{eq:minipp}) implies $f(x)$ permutes the
set of indices $A_j=\{j,j+W,j+2W, \ldots, j+(M-1)W\}$, i.e., indices
belonging to every possible window at a particular offset $j$, becomes
mapped by $f(x)$ to the set of indices $B_k=\{k,k+W,k+2W, \ldots, k+(M-1)W\}$
where 

\begin{equation}
k \equiv f_1j+f_2j^2\pmod {W}.
\label{eq:mmpp}
\end{equation}

\noindent
We conclude (\ref{eq:mmpp}) must be a permutation polynomial, otherwise
$f(x)$ would not be a permutation polynomial.

Finally, one uses induction to find that every function
 obtained by successively composing $f(x)$ (eventually generating the
 inverse function $g(x)$) generates a {\em MCF}
 interleaver.  

\end{proof}

We can observe from the previous proof that there exist {\em 
 MCF} interleavers generated by permutation polynomials
of degrees other than two. In fact, we have the following Corollary.

\begin{corollary}
Let $f(x)=\sum_{i=0}^{K}f_ix^i \pmod{N}$ be a permutation
polynomial of degree $K$. Then $f(x)$ generates a {\em MCF}
interleaver.
\label{cr:main}
\end{corollary}

\begin{proof}
The proof is identical to as in Theorem~\ref{th:main} except that
(\ref{eq:minipp}) is replaced by 

\begin{equation}
f(j+tW) \equiv \sum_{i=0}^{K}f_i j^i\pmod {W} \quad \mbox{and} \quad f(j+vW)
\equiv \sum_{i=0}^{K}f_i j^i\pmod {W},
\end{equation}

and (\ref{eq:mmpp}) is replaced by 

\begin{equation}
k \equiv \sum_{i=0}^{K}f_i j^i\pmod {W}.
\end{equation}

\end{proof}

We could had stated Corollary~\ref{cr:main} as the main theorem and 
Theorem~\ref{th:main} as a special case but we present them in this order
because the emphasis in this correspondence is on quadratic
permutation polynomials. Naturally all linear
interleavers~\cite{tak:cos:linear} (also referred 
as circular interleavers~\cite{dolinar,berrou:int} and relatively
prime (RP) interleavers~\cite{crozier1}) are {\em MCF}. However, the
error rate performance of turbo codes using linear interleavers are
constrained by the linear interleaver
asymptote~\cite{tak:cos:linear}. The almost regular permutation (ARP)
interleavers in~\cite{berrou:int} (closely related to linear
interleavers and DRP interleavers) are mentioned to have a degree of
parallel processing $pC$ dividing $N$, where $C$ is a
design parameter also dividing $N$ and $p$ any integer. However, we believe many ARP
interleavers (if not all) are {\em MCF} and therefore ARP interleavers
are stronger with respect to the 
degree of parallel processing than what is stated
in~\cite{berrou:int}. The advantage of our construction is a much
simpler description of the interleaver by a single permutation
polynomial, which we believe makes implementation simpler as
well~\cite{jpl2}. Moreover, the error performance is also not expected
to degrade against any good interleavers as shown in the following
section. 

\section{Examples and Computer Simulation Results}

We give four examples of interleavers generated by quadratic
permutation polynomials in Table~\ref{tab:examples}. The respective inverse functions are also
given for completeness and were computed using the theory
in~\cite{ryu:tak:qinv}. Because their {\em MCF} property is guaranteed
by Theorem~\ref{th:main} regardless of the choice of
the permutation polynomials, we only need to select 
permutation polynomials that yield interleavers with good error
rate performance for turbo codes. 

The interleavers in Examples 1 -- 3 were found by a
limited search for good polynomials using mainly the theory
in~\cite{sun:tak:pp}, checking for the true minimum distance
$d_{\min}$ of the associated turbo codes using the algorithm
in~\cite{gar:jsac2001} (the algorithm only finished within a reasonable
amount of time for Examples 1 and 2), and finally running computer
simulations. To the best of our knowledge, one of the most accepted indicators
for a good interleaver with respect to error performance for parallel 
concatenated turbo codes is the spread factor~\cite{crozier2,dolinar}
defined as 

\begin{equation}
D=\min_{\substack{i,j\in\{0,1,\ldots,N-1\}\\i\neq j}}\{|i-j|+|f(i)-f(j)|\}.
\label{eq:spread}
\end{equation}

\noindent
The upper bound on the spread factor
was proved in~\cite{boutillon} to be $\sqrt{2N}$ and was shown earlier~\cite{dolinar}
to be achievable or closely approximated with carefully chosen linear
interleavers. The error rate performance of 
turbo codes using any linear interleaver is constrained by the linear
interleaver asymptote~\cite{tak:cos:linear}. Therefore, the maximization of
the spread factor alone is not sufficient to guarantee a good error
performance. Nevertheless, the spread factors $D$ are computed for our
examples as a point of reference because many good constructions
attempt a maximization of the spread factor. The spread factors
obtained for Examples 1 -- 3 (i.e., the codes simulated for this
correspondence) are approximately 70\% of the upper bound $\sqrt{2N}$
independent of the degree of parallel 
processing because we use a fixed interleaver. Interestingly, the
spread factors obtained in~\cite{dobkin} are also close to 70\% of
$\sqrt{2N}$ when the degree 
of parallel processing is $M=1$ (serial processing) and with some
small decrease as the degree of parallel processing increases; the
interleavers therein found are all different for each degree of
parallel processing and the search algorithm is designed to maximize
the spread factor. 

The interleaver in Example 4, chosen by the Jet
Propulsion Laboratory, is being considered in~\cite{jpl2} 
because of its excellent performance and ease of
implementation. Example 4 is also interesting because $N=2^4\cdot 3^3
\cdot 5\cdot 7$ is composed of several different prime factors whereas
for Examples 1 -- 3, the interleaver lengths are powers of two.

\begin{table}
\centering
\caption{Examples of $MCF$ interleavers}
\begin{tabular}{|c|c|c|c|c|c|}\hline
Example & $N$ & $f(x)$ & $g(x)$ & $D$ & $d_{\min}$\\ \hline
 1  & 256 & $159x+64x^2\pmod{N}$ &  $95x+64x^2\pmod{N}$ & 16 & 27 \\
 2  & 1024 & $31x+64x^2\pmod{N}$ & $991x+64x^2\pmod{N}$ & 32 & 27 \\
 3 & 4096 & $2113x+128x^2\pmod{N}$ & $4033x+1920x^2\pmod{N}$ & 64 & - \\
 4 & 15120 & $11x+210x^2\pmod{N}$ & $14891x+210x^2\pmod{N}$ & 20 & -\\ \hline 
\end{tabular}
\label{tab:examples}
\end{table}

In all of the four examples, 16 is a factor of the interleaver length
$N$. This means that we can have a sub-block of an iterative
decoder split into 16 parallel sections without causing memory access contention
when exchanging extrinsic information with other sub-blocks.

\begin{figure}
  \centering
  \includegraphics[width=0.9\hsize]{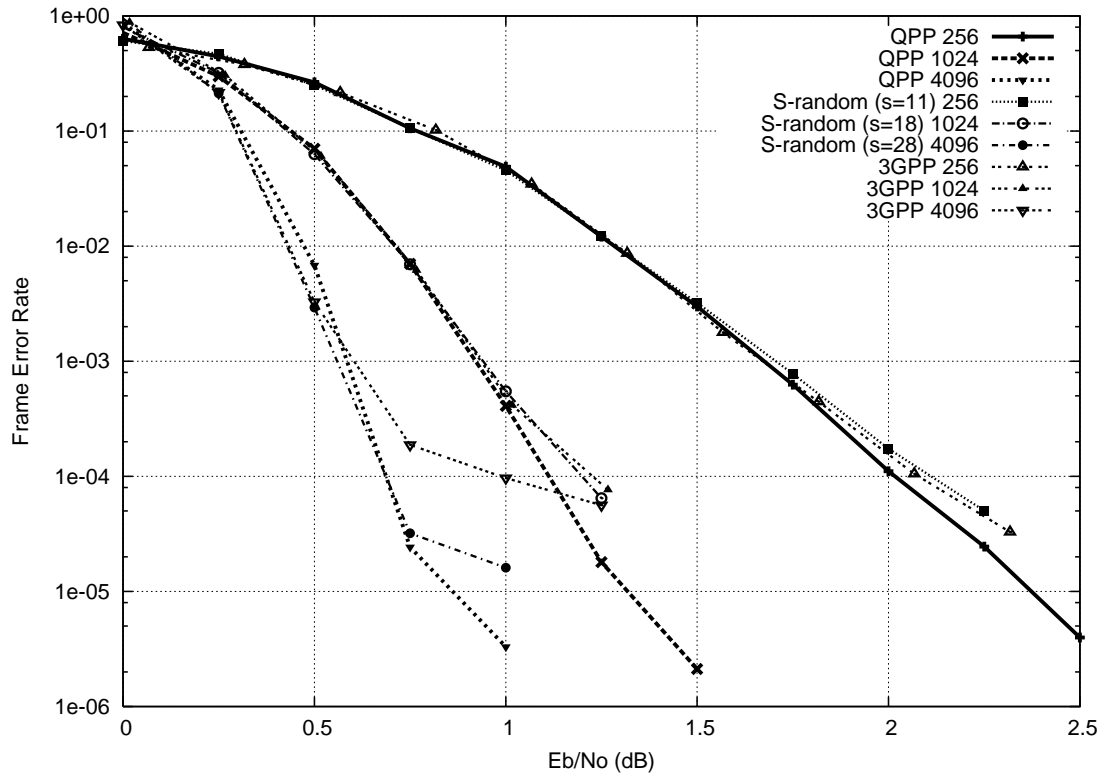}
  \caption{FER curves comparing 3GPP interleavers, S-random
    interleavers, and  our
    {\em MCF} quadratic polynomial interleavers.}
  \label{fig:pp3gpp}
\end{figure}

We now demonstrate that the restriction of an interleaver generated by
a quadratic polynomial to be {\em MCF} does not
degrade the associated turbo code error performance. On the contrary,
the quadratic interleavers generate turbo codes that have
excellent error rate performance. The simulated turbo codes are of nominal rate 1/3
for the 3rd Generation Partnership Project (3GPP) standard~\cite{3gpp}
but using {\em MCF} interleavers generated by
quadratic polynomials in Examples 1 -- 3. We use BPSK modulation and
assume an additive white Gaussian noise (AWGN) channel.  The frame error rate (FER) performance curves are
shown in Fig.~\ref{fig:pp3gpp}. We used 
eight log-MAP decoding iterations and simulated until at least 100
frame errors had been counted. The typical benchmark
S-random interleavers~\cite{dolinar} were also simulated under the same conditions. In
addition, the current 3GPP standard curves are plotted.\footnote{The
curve was obtained from~\cite{nim:isit04} but adjusted for any
termination bits rate-loss as it was done in our 
curves. The curves therein had been simulated with eight decoding
iterations and until at least 50 frame errors had been counted.} 
Additional reference curves are available in~\cite{nim:isit04}. 

It is observed from Fig.~\ref{fig:pp3gpp} that the FER performance
curves of turbo codes using the quadratic permutation polynomial
interleavers meet\footnote{The length 4096 quadratic polynomial curve
  is slightly worse for high FER's compared   with the S-random
  interleaver.} or exceed the performances of S-random interleavers
down to an FER of at least $10^{-4}$. Moreover, from the slope of the curves,
we again expect to meet or exceed the error performance against any
other interleaver down to an FER of at least $10^{-4}$. 

\section{Conclusion}

Nimbalker {\em et al.} proved that only a very small
fraction of all interleavers are
contention-free~\cite{nim:isit04}. Therefore we have shown the remarkable
fact that all permutation polynomials over integer rings
generate {\em MCF} interleavers. This property is
exceptionally important for a high-speed hardware implementation of
iterative turbo decoders because it means a potential parallel
processing of iterative decoding of turbo codes by $M$ processors for
any positive integer $M$ dividing the interleaver length
$N$. Conversely, if  one has a target of using $M$ processors, then it
suffices to choose an interleaver length $N$ which is a multiple of
$M$. We have given examples of interleavers based on quadratic
polynomials that are {\em MCF}. These interleavers
generate turbo codes with error rate performances that are expected to
meet or exceed any known interleavers for the 3GPP standard down to a
frame error rate of at least $10^{-4}$. Moreover, $MCF$ interleavers
based on quadratic permutation polynomials have virtually the simplest
generation algorithm and the least number of input parameters among all
known interleavers, which implies their very simple implementation in
software or hardware and little memory requirements.

\section*{Acknowledgment}

The author wishes to thank Dr. A. Nimbalker and Dr. D. J. Costello,
Jr., for stimulating discussions on contention-free interleavers and
for recommending practical turbo codes to be computer simulated. The
author also thanks Yen-Lun Chen for checking the contention-free
property of several interleavers and helping with a simplified proof
of Theorem~\ref{th:main}.

\bibliography{myresearch}
\bibliographystyle{IEEEbib}
\end{document}